\documentclass{article}

\usepackage[english]{babel}

\usepackage[letterpaper,top=2cm,bottom=2cm,left=3cm,right=3cm,marginparwidth=1.75cm]{geometry}

\usepackage{amsmath}
\usepackage{amssymb}
\usepackage{mathtools}
\usepackage{amsthm}
\usepackage{graphicx}
\usepackage{subfigure}
\usepackage{booktabs}
\usepackage{microtype}
\usepackage[colorlinks=true, allcolors=blue]{hyperref}
\usepackage[capitalize,noabbrev]{cleveref}
\usepackage[textsize=tiny]{todonotes}

\theoremstyle{plain}

\theoremstyle{definition}

\theoremstyle{remark}

\title{MoMoE: A Mixture of Expert Agent Model for Financial Sentiment Analysis}

\author{
Peng Shu, Junhao Chen, Zhengliang Liu, Hanqi Jiang, Yi Pan, \\
Khanh Nhu Nguyen, Zihao Wu, Huaqin Zhao, Yiwei Li, Enze Shi, ShaoChen Xu \\
\\
\textit{GyriFin Interest Group on Finance Foundation Models} \\
\\
Corresponding author: Tianming Liu (\texttt{tianming.liu@gmail.com})
}

\date{}

\begin{document}
\maketitle

\begin{abstract}
We present a novel approach called Mixture of Mixture of Expert (MoMoE) that combines the strengths of Mixture-of-Experts (MoE) architectures with collaborative multi-agent frameworks. By modifying the LLaMA 3.1 8B architecture to incorporate MoE layers in each agent of a layered collaborative structure, we create an ensemble of specialized expert agents that iteratively refine their outputs. Each agent leverages an MoE layer in its final attention block, enabling efficient task decomposition while maintaining computational feasibility. This hybrid approach creates specialized pathways through both the model architecture and the agent collaboration layers. Experimental results demonstrate significant improvements across multiple language understanding and generation benchmarks, highlighting the synergistic benefits of combining expert routing at both the neural and agent levels.
\end{abstract}

\section{Introduction}

The accurate perception and quantification of financial market sentiment have consistently represented a pivotal challenge in the domains of quantitative finance and risk management~\cite{delgadillo2024finsosent}. The ongoing digital transformation has not only exponentially increased the availability of financial text data but has also significantly augmented its complexity~\cite{liu2024large}. This complexity is reflected in multiple dimensions: the diverse structural forms of financial text, the fine-grained and often implicit nature of sentiment expression, and the immediacy with which market participants react to textual information. Despite the remarkable advancements achieved by large language models (LLMs) in the field of natural language processing (NLP), tasks specific to the financial domain—such as sentiment analysis—remain highly demanding due to the prevalence of domain-specific jargon, the subtlety of sentiment articulation, and the need to integrate cross-modal contextual information effectively~\cite{zhang2023enhancing,konstantinidis2024finllama}.

Traditional methodologies for sentiment analysis have evolved through distinct paradigms. Early lexicon-based approaches were hindered by their inability to adapt to the dynamic and context-sensitive nature of financial language, as their reliance on static dictionaries constrained their scope~\cite{mishev2020evaluation}. Machine learning models introduced data-driven paradigms but remained heavily dependent on handcrafted features curated by domain experts, thus limiting scalability and generalization~\cite{inserte2024large}. More recently, the emergence of pre-trained models such as BERT and GPT has showcased exceptional capabilities in text representation and understanding~\cite{shen2024financial}. However, their rigid architectures, computational intensity, and lack of domain-specific adaptability render them suboptimal for financial sentiment tasks, where lightweight efficiency and tailored specialization are critical~\cite{zhang2023enhancing,delgadillo2024finsosent}.

To address these challenges, this study introduces a novel sentiment analysis framework tailored to the financial domain: the \textbf{Mixture of Mixture of Experts (MoMoE)} architecture. Leveraging the LLaMA 3.1 8B foundation, the proposed framework integrates a sparsely gated mechanism with a hierarchical attention structure to enable dual-level specialization at both micro and macro scales~\cite{zhao2024aligning}. This architecture employs dynamically allocated expert modules and iterative collaborative optimization to refine sentiment representation effectively~\cite{inserte2024large}. Moreover, an enhanced load-balancing loss function is incorporated to mitigate the issue of uneven expert utilization, a common limitation of conventional MoE designs~\cite{zhang2023instruct}. This strategic combination not only achieves efficient computational resource allocation but also significantly enhances model performance, setting a new benchmark for domain-specific sentiment analysis in finance~\cite{delgadillo2024finsosent,konstantinidis2024finllama}.


Extensive experiments demonstrate that the proposed \textbf{MoMoE} architecture achieves significant advancements across multiple challenging financial sentiment analysis benchmark datasets. Specifically, the model exhibits superior robustness and accuracy in complex scenarios involving multi-entity sentiment conflicts, long-sequence dependencies, and cross-modal information integration. These results not only validate the effectiveness of the theoretical framework but also establish a new technological paradigm for intelligent decision-making systems within the financial domain.

The main contributions of this work can be summarized as follows:
\begin{itemize}
    \item A novel two-tier MoMoE architecture is proposed, which combines sparsely gated mechanisms with hierarchical attention modules to achieve dual-level specialization at both micro and macro scales.
    \item An improved load-balancing loss function is designed to address the uneven expert utilization issue commonly observed in traditional MoE architectures, thereby enhancing computational efficiency while significantly improving model performance.
    \item Extensive experiments on multiple financial sentiment analysis benchmark datasets demonstrate the superiority of the proposed method in terms of accuracy, efficiency, and generalization capability. These findings establish a new paradigm for the application of large-scale language models in the financial domain. 
\end{itemize}

\section{Related Works}

\subsection{Financial Sentiment Analysis}
Sentiment analysis has evolved significantly in financial applications, transitioning from early lexicon-based methods to machine learning approaches and more recently deep learning models that leverage large pre-trained LLMs. 

Lexicon-based methods relied on predefined word lists associated with positive or negative sentiment \cite{Pennebaker2001}, but were limited by the context-dependency of sentiment expressions \cite{Taboada2011}. The word "bull" might be associated with positive sentiment in a general lexicon, but in a financial context, a "bull market" refers to a specific market condition rather than expressing sentiment. Additionally, lexicon-based methods struggled with capturing the nuances of financial jargon, sarcasm, and the impact of negation on sentiment. Machine learning methods like SVMs \cite{Chiong2018}, Naive Bayes \cite{Kalra2019}, and Random Forests\cite{Dickinson2015} enabled capturing more complex patterns but required large labeled datasets. These methods often relied on bag-of-words or TF-IDF representations, which failed to capture the word order and context essential for understanding financial sentiment.  

The introduction of word embeddings like Word2Vec\cite{mikolov2013efficient}, GloVe\cite{pennington2014glove}, and ELMo\cite{sarzynska2021detecting} was a major milestone, allowing textual information to be represented in high-dimensional space to capture sentiment and context. Subsequent models like Doc2Vec\cite{le2014distributed} extended this to document-level embeddings. Most recently, the advent of LLMs like BERT \cite{kenton2019bert} and GPT \cite{brown2020language} has revolutionized sentiment analysis in finance \cite{su2022improving}. LLMs excel at deciphering the complexities of financial language across social media, news articles, and corporate disclosures. Their ability to process lengthy documents and even multimodal data like earnings calls enables comprehensive analysis. LLMs also show more robustness to adversarial attacks compared to prior keyword-based methods.

Studies have demonstrated the effectiveness of LLMs for sentiment analysis on social media to enhance portfolio optimization and predict stock movements\cite{steinert2023linking}. For financial news, LLMs can more accurately assign sentiment to headlines compared to prior models. Corporate disclosures like earnings calls \cite{cook2023evaluating}, press releases, and regulatory filings can also be processed by LLMs to assess sentiment and extract key insights while reducing information overload \cite{kim2024bloated}. Central bank communications like FOMC minutes and ECB policy decisions are another important application area.

The evolution of sentiment analysis techniques, culminating in powerful LLMs, has enabled significant advances in extracting sentiment from the growing volumes of financial text data to inform investment decisions and financial forecasting. Integration of LLMs with knowledge graphs and for multimodal data analysis are promising areas for further research \cite{bhatia2024fintral}. The ability of LLMs to process extensive documents, identify subtleties like sarcasm and sector-specific jargon, and exhibit enhanced resilience to adversarial attacks makes them well-suited for the challenges of financial sentiment analysis across diverse data sources \cite{curti2023let}.

\subsection{Mixture of Expert}
The MoE framework has become a critical innovation in scaling LLMs. while maintaining computational efficiency. By activating only a subset of parameters for each input token, MoE models significantly expand capacity without proportionally increasing computational costs.

Early work by \cite{shazeer2017outrageously} introduced the Sparsely Gated MoE model, demonstrating that routing input tokens to different expert networks could improve parameter efficiency in deep learning architectures. This approach provided an effective way to increase model size without a linear growth in FLOPs.
Building on this foundation, \cite{lepikhin2020gshard} proposed GShard, which applied MoE to large-scale machine translation models and introduced more efficient routing mechanisms.Their work showed that MoE-based transformers could scale significantly while keeping computational costs manageable.
Expanding on MoE efficiency, \cite{fedus2022switch} developed the Switch Transformer, modifying previous architectures by routing each token to a single expert rather than multiple. This refinement reduced computational overhead and improved model stability during training.

Further advancements came with \cite{zoph2022designing}, who integrated sparse expert activation into Google’s Pathways Language Model (PaLM), allowing trillion-parameter scaling while preserving strong few-shot learning capabilities. Their findings underscored the adaptability of MoE architectures in large-scale LLMs.

More recently, \cite{team2024gemini} explored hybrid MoE architectures in Gemini 1.5, combining dense and sparse expert layers to enhance multi-modal reasoning and generalization across NLP tasks. In parallel, \cite{du2022glam} examined MoE training instabilities in GLaM (Gated Latent MoE) and introduced optimizations to balance computational loads and improve gradient updates, addressing key challenges in training stability.

These developments have established MoE as a key tool for scaling LLMs efficiently, with ongoing research focusing on improving routing strategies, load balancing, and inference performance.

\section{Methodology}

\subsection{MoE Layer}
We construct our financial sentiment analysis model based on the LLaMA 3.1 8B architecture. We refer to this model as LlaMoE. Since our focus is on a specific downstream task, we modify the feed-forward network (FFN) layer in the final attention block, replacing it with a custom-designed MoE layer. The rationale for this modification lies in the observation that the final attention layer extracts higher-level semantic features \cite{geva2020transformer}. We hypothesize that this minor adjustment is sufficient to enhance performance on our financial sentiment analysis task while maintaining computational efficiency. Additionally, this approach enables us to fine-tune the LLaMA 3.1 8B model using limited computational resources and data.

\begin{figure}[ht]
\begin{center}
\centerline{\includegraphics[width=0.5\textwidth]{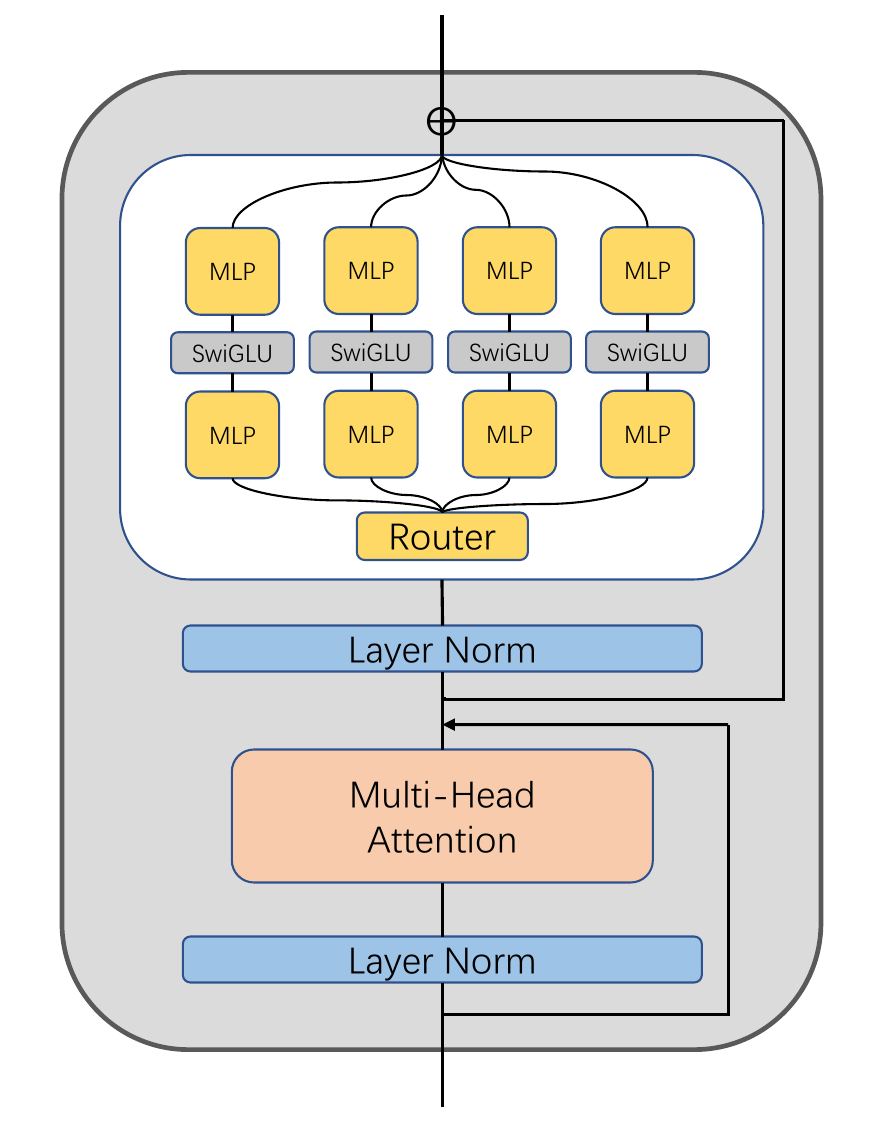}}
\caption{Overall architecture of our LLaMoE model. We replace the FFN in the final attention block with a MoE layer. The router dynamically selects the top-k experts, each consisting of a three-layer structure incorporating SwiGLU activations. The MoE output is then combined with the residual connection from the preceding Multi-Head Attention layer to ensure stable gradient flow and information retention.}
\label{model_structure}
\end{center}
\end{figure}

The designed MoE layer is shown in figure \ref{model_structure}. Specifically, we replace the FFN layer in last attention layer to be a MoE layer. Our MoE layer consists of 4 layers:
\begin{itemize}
    \item A MLP router
    \item Two MLP layers
    \item  A SwiGLU activation function \cite{riquelme2021scaling} between two MLP layers
\end{itemize}
Given the outputs from last Multi-Head Attention block,Our MoE model applies a single MLP layer to compute the gate score \( g \in \mathbb{R}^{T \times E} \), where \( T = B \times L \) represents the total number of tokens, with \( B \) denoting the batch size and \( L \) the sequence length. \(E\) is the number of experts. The gate scores are transformed into routing probabilities \(r_{t,e} \in \mathbb{R}^{T \times E}\)using the softmax function:
\[r_{t,e} = \text{softmax}(g)\]
The model then selects the top \(k\) experts with the highest gate scores for each token, ensuring that only a subset of experts is activated per forward pass. This selective routing mechanism allows the MoE structure to maintain computational efficiency by leveraging sparsity, as only a fraction of the total expert parameters are utilized at any given time. Once the selected experts process the token representations independently, their outputs are aggregated based on the computed gate scores, ensuring a weighted contribution from each activated expert. The final output of the MoE layer is then combined with the residual connection from the output of the attention block, facilitating stable gradient flow and preserving essential contextual information. This aggregated representation serves as the final output of the last attention layer.

To address the imbalance in expert utilization, we incorporate an auxiliary load balancing loss, as proposed in \cite{shazeer2017sparsely, lepikhin2020gshard}. This additional loss term encourages the model to distribute token assignments more evenly across all experts, mitigating the risk of certain experts being overutilized while others remain underutilized. Without such a mechanism, the model may develop a preference for a small subset of experts, leading to inefficient parameter utilization and potential bottlenecks in training. The load balancing loss is mathematically defined as:
\[
\mathcal{L}_{\text{balance}} = \frac{1}{E} \sum_{e=1}^E f_e \cdot p_e
\]

where
\[
f_e = \frac{\sum_{t=1}^T \mathbb{I}(e \in \text{TopK}(r_t))}{T}.
\]

\[
\mathbb{I}(e \in \text{TopK}(r_t)) =
\begin{cases} 
1, & \text{if } e \in \text{TopK}(r_t), \\
0, & \text{otherwise}.
\end{cases}
\]

\[
p_e = \frac{\sum_{t=1}^T r_{t,e}}{T}.
\]

\(f_e\) is the fraction of tokens routed to expert \(e\) and \(\mathbb{I}(e \in \text{TopK}(r_t))\) is an indicator function. This ensures that all experts' utilization is considered during training.
\(p_e\) is the average routing probability for each expert \(e\). 
The final training objective combines the task-specific loss \(\mathcal{L}_{\text{train}}\) with the load balancing loss \(\mathcal{L}_{\text{balance}}\)
\[
\mathcal{L} = \mathcal{L}_{\text{train}} + \alpha \mathcal{L}_{\text{balance}}
\]
where \(\alpha\) is a hyperparameter controlling the weight of the balance loss.

\subsection{Supervised Fine-Tuning}
To train our modified LLaMoE model, we employ SFT on financial sentiment analysis data by optimizing the total training loss. Specifically, we construct the training data by formatting financial news along with their corresponding sentiment labels as natural language sequences. We generated 15 questions and 10 prefixes for preceding answer for the training data to ensure the robustness in sentiment classification. Appendix \ref{appendix_1} shows more details. This formulation enables the model to learn sentiment prediction in a generative manner, aligning with the autoregressive nature of the transformer architecture. During fine-tuning, we kept the entire model frozen except for the proposed MoE layer. The training objective follows the next-token prediction paradigm, where the model learns to generate the next token based on the preceding context. This approach ensures that the model effectively captures both financial text patterns and sentiment-relevant features, ultimately enhancing its ability to generate sentiment-aware predictions.


\subsection{MoMoE}

\begin{figure}[ht]
\vskip 0.2in
\begin{center}
\centerline{\includegraphics[width=0.6\textwidth]{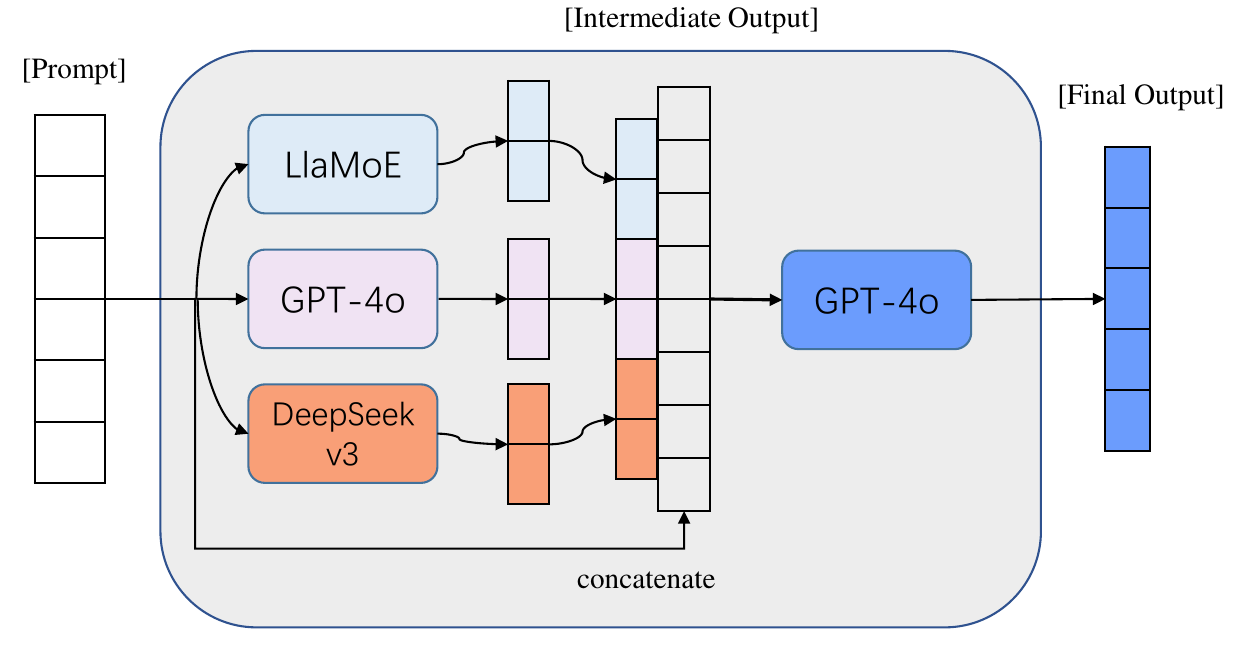}}
\caption{Illustration of our single-layer agent system following the MoA structure. The input prompt is processed independently by three agents: LLaMoE, GPT-4o, and DeepSeek V3. Their intermediate outputs are concatenated with the original prompt and subsequently fed into a final decision-making agent (GPT-4o) to produce the final classification output.}
\label{momoe}
\end{center}
\vskip -0.2in
\end{figure}

To further enhance the performance of our Supervised Fine-Tuned LLaMoE model, we incorporate the Mixture of Agents (MoA) structure as introduced in \cite{wang2024mixture}. This leads to the development of a novel model architecture, which we refer to as Mixture of Mixture of Experts (MoMoE). The MoMoE architecture builds upon the MoA framework by combining multiple agent-based models to process input data in parallel, thereby improving decision-making capacity and model flexibility. In our implementation, we apply a single layer of MoA to augment the LLaMA-MoE by directing the same input prompt to three distinct agents: GPT-4o, DeepSeek V3, and the original LLaMoE model. The outputs of these three agents are concatenated along with the original input prompt and passed to a final decision-making agent—GPT-4o in this case—which produces the final classification output. The general structure of the MoMoE model is illustrated in Figure \ref{momoe}, showing the flow of information across agents and the final aggregation mechanism. 

\section{Experiment Results}
\subsection{Dataset}
We utilized several prominent datasets to train and evaluate our financial sentiment analysis models:

\subsubsection{FinancialPhraseBank v1.0}
The FinancialPhraseBank v1.0 dataset \cite{malo2014good}, obtained from Hugging Face, consists of 4,840 sentences extracted from English-language financial news articles. Each sentence is labeled as positive, negative, or neutral, with annotations provided by 16 financial experts to ensure high-quality sentiment classification. The dataset offers subsets based on annotator agreement levels ranging from 50\% to 100\%, allowing for flexibility in model training and evaluation. This dataset enables the development of models that can accurately classify sentiment at the sentence level in financial news.

\subsubsection{SEntFiN 1.0 }
SEntFiN 1.0, sourced from arXiv \cite{sinha2022sentfin}, is a comprehensive dataset containing 10,753 news headlines with entity-specific sentiment annotations. Notably, 2,847 headlines include multiple entities, often with conflicting sentiments. This dataset provides a nuanced resource for fine-grained sentiment analysis, supporting the development of models capable of discerning sentiment at the entity level within complex financial news contexts. By leveraging SEntFiN 1.0, we can train models that capture the intricacies of sentiment expression in financial headlines.

\subsubsection{Kaggle Sentiment Analysis for Financial News}
The Kaggle Sentiment Analysis for Financial News dataset \cite{sentimentanalysisfinancialnews} comprises financial news headlines labeled with sentiments from the perspective of retail investors. Each entry includes a headline and its corresponding sentiment—positive, negative, or neutral. This dataset facilitates the training of models focused on headline-level sentiment detection, enabling us to develop solutions tailored to the needs of individual investors who rely on financial news sentiment for decision-making.

\subsubsection{Kaggle Twitter Financial News Sentiment Dataset}
Obtained from Hugging Face, the Kaggle Twitter Financial News Sentiment Dataset is an English-language dataset containing an annotated corpus of finance-related tweets\cite{twitterfinancialnews}. The dataset is used to classify the sentiment of these tweets, supporting the development of models that analyze sentiment in financial social media content. By incorporating this dataset into our experiments, we can extend our sentiment analysis capabilities to the domain of financial discussions on social media platforms.

\subsubsection{FiQA}
The FiQA dataset \cite{de2018inf}, specifically the INF-UFG subset, provides a valuable resource for evaluating models on sentiment and aspect prediction tasks using entity- and aspect-specific annotations of financial tweets and news headlines. This dataset challenges models to handle complex financial expressions and diverse sentiment-bearing contexts, contributing significantly to the field of financial sentiment analysis. By incorporating the FiQA dataset into our experiments, we can assess the performance of our models in capturing fine-grained sentiments and aspects, further enhancing their applicability in real-world financial scenarios.

These diverse datasets were instrumental in training and evaluating our sentiment analysis models, providing rich sources of financial textual data across various contexts, including news articles, headlines, social media posts, and entity-specific annotations. By leveraging these datasets, we can develop robust models capable of accurately capturing sentiment in a wide range of financial text, enabling more informed decision-making for investors and financial professionals.

\subsection{Experiment Settings}
In our LLaMoE model, we implement 4 experts and select the top 2 experts with the highest gate scores for activation. Each expert's MLP module has a hidden dimension set to 4 times the hidden size of the Multi-Head Attention output. For the auxiliary loss, we use a weighting factor of \( \alpha = 0.01 \). 

We train our model on 8 NVIDIA A5000 GPUs, each with 24GB of memory. The dataset is split into 90\% for training, while 10\% is reserved for validation, from which we sample 200 data pairs as the test set. The model is trained for 1 epoch with a batch size of 64 and a learning rate of \( \eta = 1 \times 10^{-4} \).

\subsection{Results}

\begin{table*}[t]
\caption{Experiment results on the financial sentiment analysis classification task. Our MoMoE model achieves state-of-the-art performance across all evaluation metrics.}
\label{results}
\vskip 0.15in
\centering
\begin{small}
\begin{sc}
\begin{tabular}{lcccc}
\toprule
Model & Accuracy & F1-score & Precision & Recall \\
\midrule
FinBERT              & 73.0\%  & 72.9\%  & 73.8\%  & 73.0\%  \\
DeepSeek V3          & 75.0\%  & 74.9\%  & 76.0\%  & 75.0\%  \\
GPT-4o               & 74.8\%  & 74.3\%  & 77.4\%  & 74.5\%  \\
LLaMoE               & 75.5\%  & 74.7\%  & 75.9\%  & 75.5\%  \\
MoMoE (LLaMoE + MoA) & \textbf{75.5\%}  & \textbf{76.6\%}  & \textbf{78.7\%}  & \textbf{75.5\%}  \\
\bottomrule
\end{tabular}
\end{sc}
\end{small}
\vskip -0.1in
\end{table*}

\begin{figure*}[ht]
\vskip 0.2in
\begin{center}
\centerline{\includegraphics[width=1\textwidth]{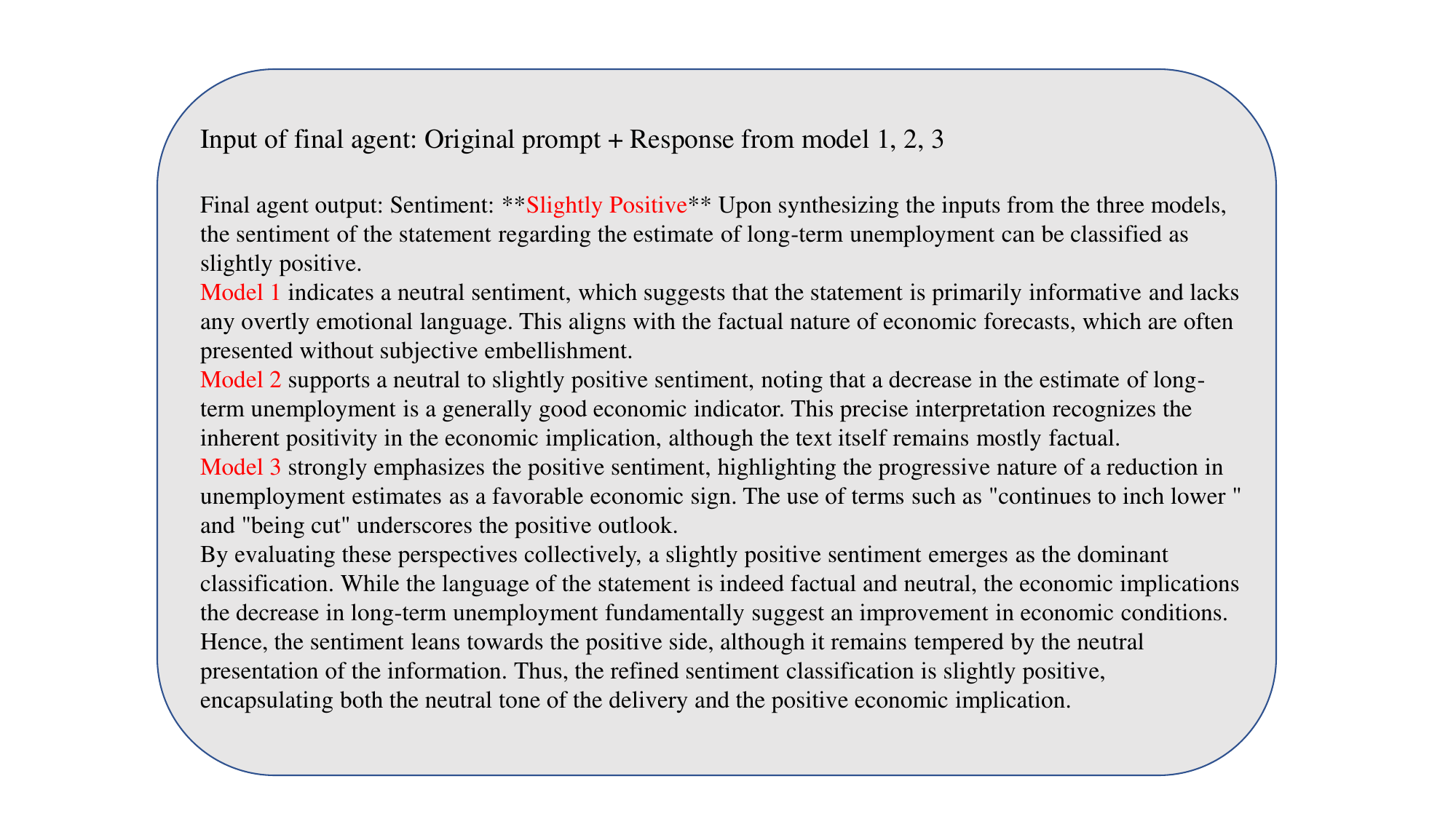}}
\caption{One example illustrates how the final agent overrides an incorrect intermediate prediction and determines the correct classification outcome.}
\label{output_example}
\end{center}
\vskip -0.2in
\end{figure*}

To assess our model's performance, we compare it against state-of-the-art LLMs, including GPT-4o and DeepSeek V3. We also evaluate FinBERT \cite{araci2019finbert}, a transformer model specialized in financial text analysis. Table~\ref{results} reports results across four key metrics: Accuracy, F1-score, Precision, and Recall.

\textbf{MoMoE} (LLaMoE + MoA) achieves the best overall performance, with an F1-score of \textbf{76.6\%} and a Precision of \textbf{78.7\%}, outperforming all baselines. The combination of MoE and MoA improves classification quality by selecting specialized experts while incorporating multiple decision-making agents.

We observe that in MoMoE, the final agent evaluates the outputs of the three intermediate models, analyzing their responses and reasoning about the underlying perspectives that led to each decision. It then integrates these outputs, considering both the original prompt and the intermediate predictions, to form the final classification decision. 

In many cases, this multi-agent collaboration helps correct errors from individual models. When one or more intermediate agents produce incorrect predictions, the final agent can still infer the correct classification label by leveraging information from the more reliable outputs. Figure \ref{output_example} presents an example where MoMoE successfully mitigates errors from intermediate agents and generates an accurate sentiment classification.

However, in some instances, the final agent may be influenced by a majority of intermediate models producing the same incorrect output. When two agents arrive at the same but incorrect conclusion, their combined weight can bias the final decision, leading to misclassification. Addressing this limitation is an important direction for future work, where we aim to improve the robustness of the final agent by enhancing its reasoning ability and introducing mechanisms to better evaluate conflicting outputs among agents.

LLaMoE, our MoE-based variant, also surpasses the baselines, reaching \textbf{75.5\%} Accuracy and an F1-score of \textbf{74.7\%}. The results suggest that expert routing improves specialization, enabling the model to capture different aspects of financial sentiment. MoMoE further refines this approach by integrating multiple agent responses, contributing to better overall consistency.

Among the baseline models, DeepSeek V3 performs the best, with an F1-score of \textbf{74.9\%} and an Accuracy of \textbf{75.0\%}, surpassing GPT-4o and FinBERT. DeepSeek V3 likely benefits from large-scale pretraining, allowing it to recognize financial language patterns more effectively. GPT-4o shows strong Precision (\textbf{77.4\%}) but a lower F1-score (\textbf{74.3\%}), indicating an imbalance between Precision and Recall.

FinBERT, despite being tailored for financial text, records the lowest Accuracy (\textbf{73.0\%}) and F1-score (\textbf{72.9\%}). This suggests that domain-specific pretraining alone is insufficient to compete with larger, more flexible models trained on broader datasets.

Overall, MoMoE achieves the highest F1-score and Precision, making it the most reliable model in this setting. The results indicate that expert routing and multi-agent collaboration improve financial sentiment classification by allowing models to focus on relevant features and refine predictions through multiple perspectives.

\section{Disscussion}
In this paper, we present MoMoE, a novel Mixture of Mixture of Experts framework that extends the LLaMA 3.1 8B model by integrating a Mixture of Experts mechanism, referred to as LLaMoE. Additionally, we incorporate the Mixture of Agents framework to further refine classification accuracy by leveraging multiple agent-based predictions. 

Through supervised fine-tuning, MoMoE demonstrates superior performance on financial sentiment analysis, achieving state-of-the-art results compared to leading large language models, including GPT-4o and DeepSeek V3. Furthermore, MoMoE surpasses FinBERT, a model specifically designed for financial text analysis, highlighting the advantages of expert routing and multi-agent decision-making over static transformer architectures. 

Our results indicate that MoMoE effectively captures the nuances of financial sentiment by dynamically selecting specialized experts while incorporating multiple perspectives for classification. This multi-expert and multi-agent approach enhances both precision and generalization, leading to more reliable predictions. 

Despite its strong performance, MoMoE is not without limitations. In some cases, the final agent may be influenced by dominant incorrect intermediate outputs, leading to misclassification. Future work will focus on improving the robustness of the agent-based reasoning process by introducing adaptive weighting mechanisms and confidence-aware aggregation strategies. Additionally, we aim to explore scaling MoMoE to larger architectures and extending its applicability to broader financial NLP tasks, including event-driven sentiment analysis and market prediction.

Overall, our study demonstrates the effectiveness of combining MoE and MoA architectures in financial sentiment classification, setting a strong foundation for further research in expert-guided large language models.

\bibliographystyle{plain}
\bibliography{moe_reference}

\newpage
\appendix
\section{Question and Prefix Templates for SFT}
\label{appendix_1}
Here, we present the set of questions and prefix templates used before the sentiment labels. We utilize GPT-4o to generate 15 candidate sentences for the question component and 10 variations for the prefix preceding the answer. These templates help diversify model inputs and ensure robustness in sentiment classification.

Questions:
\begin{itemize}
    \item Can you analyze this financial sentiment?
    \item Please evaluate the sentiment of the following text:
    \item Determine the sentiment of this financial statement.
    \item Assess whether this text is positive, negative, or neutral.
    \item Here is a financial text for sentiment analysis:
    \item Classify the sentiment of this passage:
    \item Identify the tone of the following financial text:
    \item Evaluate this statement for sentiment polarity.
    \item What is the sentiment of this financial report?
    \item Is this sentiment optimistic, pessimistic, or neutral?
    \item Analyze the sentiment of the following:
    \item Assess the mood of this financial description.
    \item What sentiment does this statement convey?
    \item Classify the financial sentiment in the text below.
    \item Analyze the sentiment expressed in this passage.
\end{itemize}

Prefix for preceding the answer:
\begin{itemize}
    \item The sentiment of this text is:
    \item This passage conveys a sentiment of:
    \item Analyze and determine the sentiment as:
    \item The tone of the statement is:
    \item Classify the following text's sentiment as:
    \item The correct sentiment is:
    \item This statement reflects a sentiment of:
    \item The mood expressed in this text is:
    \item Determine the sentiment conveyed as:
    \item The following text exhibits a sentiment of:
\end{itemize}

\end{document}